\documentclass[sigconf,nonacm]{acmart}
\pdfoutput=1

\usepackage{hyperref}
\hypersetup{pdfauthor={François Labrèche},pdftitle={Shedding Light on the Targeted Victim Profiles of Malicious Downloaders}}
\usepackage{multirow}
\usepackage{subfig}
\usepackage{url}
\PassOptionsToPackage{hyphens}{url}

\newtheorem{definition}{Definition}

\begin{document}

\title{Shedding Light on the Targeted Victim Profiles of Malicious Downloaders}

\author{Fran\c{c}ois Labr\`eche}
\affiliation{%
  \institution{\'Ecole Polytechnique de Montr\'eal}
  \streetaddress{2500 Chemin de Polytechnique}
  \city{Montr\'eal}
  \country{Canada}
  \postcode{H3T 1J4}}
\email{francois.labreche@polymtl.ca}

\author{Enrico Mariconti}
\affiliation{%
  \institution{University College London}
  \streetaddress{35 Tavistock Square}
  \city{London}
  \country{United Kingdom}
  \postcode{WC1H 9EZ}}
\email{e.mariconti@ucl.ac.uk}

\author{Gianluca Stringhini}
\affiliation{%
  \institution{Boston University}
  \city{Boston}
  \country{United States}
  \postcode{02215}}
\email{gian@bu.edu}

\begin{abstract}
  Malware affects millions of users worldwide, impacting the daily lives of many people as well as businesses. Malware infections are increasing in complexity and unfold over a number of stages. A malicious downloader often acts as the starting point as it fingerprints the victim's machine and downloads one or more additional malware payloads. Although previous research was conducted on these malicious downloaders and their Pay-Per-Install networks, limited work has investigated how the profile of the victim machine, e.g., its characteristics and software configuration, affect the targeting choice of cybercriminals.

In this paper, we operate a large-scale investigation of the relation between the machine profile and the payload downloaded by droppers, through 151,189 executions of malware downloaders over a period of 12 months. We build a fully automated framework which uses Virtual Machines (VMs) in sandboxes to build custom user and machine profiles to test our malicious samples. We then use changepoint analysis to model the behavior of different downloader families, and perform analyses of variance (ANOVA) on the ratio of infections per profile. With this, we identify which machine profile is targeted by cybercriminals at different points in time.

Our results show that a number of downloaders present different behaviors depending on a number of features of a machine. Notably, a higher number of infections for specific malware families were observed when using different browser profiles, keyboard layouts and operating systems, while one keyboard layout obtained fewer infections of a specific malware family.

Our findings bring light to the importance of the features of a machine running malicious downloader software, particularly for malware research.
\end{abstract}


\maketitle

\section{Introduction}
Malicious software, i.e., malware, infects tens of thousands of machines every day around the globe~\cite{kaspersky}.

Once a user has visited a malicious link and has been redirected, an exploit kit or the website's code will infect them with a malicious software. This malicious software can take many forms, one of them being a downloader software. In this scenario, its sole purpose will be to download other malicious software. 

\begin{figure}[htp]
\centering
 \includegraphics[width=0.7\columnwidth]{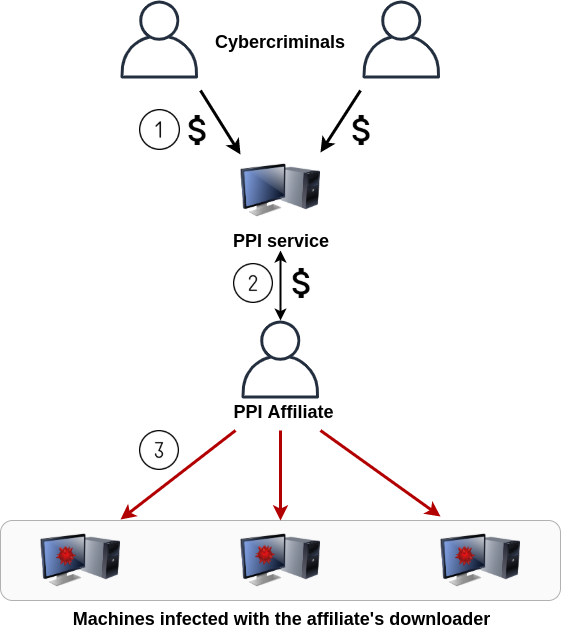}
  \caption[The Pay-Per-Install distribution model]{The Pay-Per-Install distribution model. 1) Cybercriminals pay a PPI service to install their malware on machines. 2) A PPI affiliate obtains malware to install on victim machines through the PPI service. 3) The PPI affiliate installs malware payloads on victims infected with its malicious downloader.}
 \label{fig:ppi}
 \vspace*{-2mm}
\end{figure}

For example, a malicious actor will infect several machines with their downloader, and then sell the access to one or more other criminals, who will install spam bots, information stealers, ransomware, etc., on the machine. This process is part of the Pay-Per-Install (PPI) distribution model, where the downloader's author will sell access to the infected machine to another malicious actor through the PPI service. Figure~\ref{fig:ppi} illustrates the PPI distribution model. These transactions are made privately, and information regarding these transactions, the actors involved, and the platforms on which they occur is difficult to obtain. 

Particularly, little is known on how cybercriminals select the malicious file(s) to send to the victim through the downloader, i.e., how much, if any, fingerprinting is done on the victim and their machine, to decide which malware to infect them with. 

Researchers have studied malware in multiple previous research, while downloader software used for malware is often overlooked in research. Some works have studied downloader software, mainly through a study of PPI models~\cite{caballero2011measuring, kotzias2016measuring, thomas2016investigating}, without a large focus on what impact the various features of the victim machine has on the PPI network customer's targeting choice.

In this paper, we will use a machine's features to test whether the machine is targeted by one or more cybercriminals, by executing various families of downloaders daily and identifying downloaded malicious payloads, over a one-year period. 

With our approach, we aim at reverse engineering the targeting choices of criminals. However, with respect to classic reverse engineering techniques that analyze the code, we deduce what influenced the decision-making process of cybercriminals by simultaneously running multiple VMs with different configurations (profiles). This approach is more time effective and will not be influenced by the complexity and obfuscation of the downloader's code. 

We will achieve this by executing a downloader family's samples over a large set of Virtual Machines (VMs) with different profiles, in order to establish which VM gets infected with which family of malware. Thus, our hypothesis is the following: the malicious software sent to a victim through a downloader will vary according to the machine's features.

While a downloader and malware can be combined in a single malware, we will, in this work, focus on analyzing single downloaders. In order to verify our hypothesis, our research will aim at:
\begin{enumerate}
	\item Establishing an automated sandboxed testing environment for downloaders.
	\item Establishing the correlation between the profile of the victim machine and the downloaded malicious payload(s). 
\end{enumerate}

Although previous research has analyzed user profiles more at risk through clinical trials~\cite{lalonde2013clinical}, the correlation between the user and the payload requested by a downloader has not been studied in detail. 

In summary, this paper makes the following contributions:
\begin{enumerate}
	\item We establish a fully automated sandboxed testing environment, capable of running malicious downloader executables on virtual machines configured to match specific machine and user profiles.
	\item We demonstrate what cybercriminals target through PPI networks by observing how a number of downloaders behave differently depending on the profile of the VM running the executable. Our results highlight an existing link between the malicious actors' downloaders and the browser session, the keyboard layout and the display language of machines.
\end{enumerate}

\section{Related Work}
Malicious downloaders, or droppers, are often observed when analyzing malware. Although the downloader is often paired with the malware, researchers focus more on the final malicious payload. We present, in the following section, recent research that analyze downloaders and their PPI networks. 

\subsection{Downloader Families and Samples Analysis}
Some researchers have analyzed specific downloaders and large campaigns in detail.

Rossow \emph{et al.}~\cite{rossow2012large} observed a large number of malware downloaders from 23 families between February 2010 and February 2012 and identified their properties and behavior. They identified the means of communication they use to reach their command and control (C\&C) server or other infected machines, by reassembling and parsing numerous carrier protocols. 
 
They observed these samples over a period of time and identified how frequently the domains and infrastructure change, and how long the downloader remains active. Their analysis showed that 48\% of downloaders actively operated for more than a year. They then inspected how downloaders request their malicious software to install on the machine, and recreated it to farm samples. 

They observed the number of executables served, and established that polymorphism was used by 8 of the 9 families of malware gathered. 

Kwon \emph{et al.}~\cite{kwon2015dropper} approached the analysis of downloaders by creating influence graphs of downloaders, and identifying differences between malicious and benign graphs. 

They extracted a total of 19 million influence graphs from their dataset.  An analysis of these graphs revealed that 22.4\% (15,115) of downloaders have a valid digital signature, influence graphs with a large diameter are mostly malicious, influence graphs with slow growth rates are mostly malicious, and malware tend to download fewer files per domain. They then used this graph representation of downloaders to extract several features. Using these, they employed supervised learning to create a detection model and then identified the most important features linked to malicious downloaders. 

\subsection{Pay-per-Install} 
Downloader software often has a presence in PPI networks. This software is used by PPI providers to install their clients' malware onto compromised machines.

Caballero \emph{et al.}~\cite{caballero2011measuring} infiltrated four PPI networks and ran their downloaders in a closed environment, from August 2010 to February 2011. They harvested over a million client executables using servers across 15 countries. They observed over time the different malware samples downloaded by the PPI networks and clustered them into their respective families and types, using their network activity. They also observed the repacking rate of malware. The top 10 families show that they are repacked, i.e., their code is re-obfuscated, every 6.5 day on average. They observed samples running anti-VM techniques; some samples even removed or added them while in a campaign, without any apparent reason. Another noteworthy observation was that a number of executables extracted from the downloaders are in fact other PPI downloaders, hinting at the fact that there might be arbitrage in the PPI market, i.e., that PPI services buy and resell their services between themselves to make a profit on varying prices. The downloaded samples also differed depending on the location of the machine, and an analysis of PPI forums showed that the price varies according to the location of the purchase of installs.

Kotzias \emph{et al.}~\cite{kotzias2016measuring} analyzed potentially unwanted programs (PUP), which consist in software that is approved by the user, knowingly or not, but exhibit a behavior detrimental to them. 

Their first step was to identify top PUP publishers by their signed software name. They then clustered publishers by running a name similarity algorithm, among other techniques. They looked at the prevalence of PUP, and 54\% of the dataset hosts had some form of PUP installed. Compared to legitimate software, the top PUP enterprise ranked 15th, which shows how widespread these PUPs are. They then established a relation graph between installers to see which installs which, and then identified PPI services by their high count of outgoing relations and ingoing relations, which suggests they sell installs. 

They also found that the majority of PUP are installed by other PUP. In total, they observed 71 PUP publishers that clustered to malware. However, this number is small in contrast to their total number of publishers.

Thomas \emph{et al.}~\cite{thomas2016investigating} explored the ecosystem of PPI services to establish what adware they distribute. They explored the 4 largest PPIs from their network. They established an infrastructure to collect software distributed by these PPIs on a regular basis. They installed downloaders from these PPIs, and observed that these downloaders sent to their server the client's OS and service pack version, the Web browsers and their version, and potentially unique identifiers including a MAC address. The server then sent 5 to 50 potential offers to the downloader. In this work, they then implemented \emph{milkers} and collected 446,852 offers through this.

They monitored the price of these offers through forums and various websites advertising these online. They clustered the results into families based on a multitude of characteristics. 

Additionally, they analyzed the length of campaigns by looking at offers from the PPI networks. Through the forums advertising these, they found that the prices differ per country. Finally, they identified the lifetime of adware sent through PPIs, which varied from 0.75 hours to the entire monitoring window of 220 days.

Finally, Kwon~\emph{et al.}~\cite{kwon2016catching} implemented Beewolf, a software built to identify malicious campaigns by using unsupervised learning to identify the \emph{locksteps} of malicious downloaders retrieving their payloads. This work analyzes the global behavior of downloaders in order to improve the detection of malicious campaigns, and not necessarily single malicious files. It identifies overlaps between malware delivery campaigns and PUP delivery campaigns.

\subsection{Summary}
While many projects have focused on improving the detection of malicious software, exploit kits and malicious websites, downloaders have seen less focus from the scientific community. 

Some of these works~\cite{rahbarinia2016real,ife2019waves} have focused on analyzing malware downloads, which includes the analysis of downloaders fetching malicious payloads, although they do not focus on malicious downloaders behavior.

Researchers have analyzed large-scale downloader campaigns through adware, i.e., PUP, and, although specific downloaders and PPI networks have been studied in detail, no research has shown the impact the user and its machine configuration have on the downloaded malware by the downloader. More notably, no testing framework has implemented the use of VMs that modify their configuration while testing downloader samples.

\section{Methodology}
Our approach consists in building an automated sandboxed environment, containing multiple VMs, where malicious downloader samples of various families are run with multiple machine profiles. 

\begin{definition}
A \emph{machine profile} consists of a machine $M$ possessing $n$ associated features $M_f = \{M_1, M_2, .., M_n\}$.
\end{definition}

A sample is automatically tested on multiple machine profiles in order to gather what executable(s) is downloaded depending on which profile was used. An analysis of the time series of infections for every feature and payload family is then performed, in order to assess what profile is targeted by the actors behind malicious downloaders. 

\begin{definition}
A \emph{feature} $F$ of a machine $M$ can be summarized as a modifiable piece of software or hardware with a subset of $l$ values, such that $F = \{F_1, F_2, .., F_l\}$.
\end{definition}

\subsection{Testing Environment}
Our first research objective is to build an automatic testing environment to run downloaders using various machine profiles. 

The design of our platform is inspired from previous research~\cite{thomas2016investigating}, and uses the Cuckoo Sandbox~\cite{sandbox2013automated} framework to execute samples. 

The Cuckoo framework provided us with the data necessary for our analysis, and ended up successfully capturing malware detonations.

\begin{figure}[tbp]
\centering
 \includegraphics[width=0.6\columnwidth]{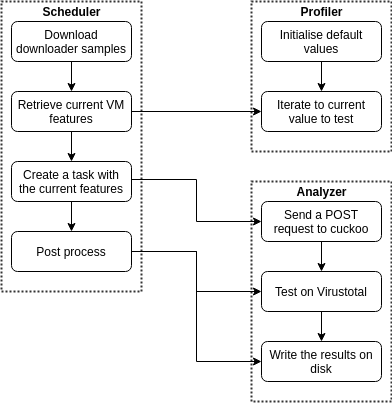}
  \caption[Our automation framework.]{Our framework depicting our downloader testing environment}
 \label{fig:framework}
 \vspace*{-6mm}
\end{figure}

This testing environment consists of a cluster of machines automatically launching VMs in a sandbox, according to a 1) Scheduler, while using a predefined user profile for each machine through a 2) Profiler. Each VM is provided with a downloader sample to run, and data is retrieved from the execution and then compiled with the 3) Analyzer, where the downloaded files' malware families are also identified. Our framework is depicted in Figure~\ref{fig:framework}. Here we detail the different sections:

\textbf{Scheduler}: This module receives the encrypted downloader sample through a secure stream, asks the profiler to build a VM, and then decrypts the sample and launches it in the VM.

\textbf{Profiler}: This module is responsible for following the experimental design, in terms of features to test. In this module, a profile will be built according to the current features in the queue, and sent back to the scheduler. It will also specify the exit node, i.e., the country, to use through a virtual private network (VPN).

\textbf{Analyzer}: This module is responsible for gathering data on the execution of the downloader samples. It uses the Cuckoo sandbox to capture network traces, process information and identify any downloaded files. This module tests any downloaded file by the downloader on VirusTotal~\cite{total2012virustotal}, a platform that tests files against multiple antivirus software. The module then establishes the file's family according to the naming of security vendors.

\subsection{Automation Setup}
Our physical testing environment consists of a cluster of 10 servers, each one containing an Intel Xeon E5405 2.00 GHz processor and 8 GB memory. One server is used as the experiment manager, where Cuckoo Distributed and the VPNs are installed. This machine dispatches the tasks of our scheduler to the other servers' Cuckoo frameworks through Cuckoo Distributed. One network interface is used to access the servers, while a separate network interface is used for each of the VPNs, in order to isolate the Internet access of our VMs. Every VM has Internet access only through its assigned VPN by the scheduler. OpenVPN\footnote{https://openvpn.net/} is used to install and manage the VPN connections. Our setup is fully automated: xCAT\footnote{https://xcat.org/} is used to automate the installation of Ubuntu LTS on our servers, and Ansible\footnote{https://www.ansible.com/} scripts are employed to configure our setup, which consists of the following steps: 
\begin{enumerate}
	\item Install and configure libvirt with KVM for the virtualization
	\item Copy images of Windows XP, Windows 7 and  Windows 10 on each server
	\item Create the VMs and their default snapshots
	\item Setup 10 VPNs
	\item Install and configure the Cuckoo and Cuckoo Distributed frameworks
	\item Configure our virtualization software according to Cuckoo's requirements
\end{enumerate} 

Each server has three VMs, and our cluster totals 27 VMs which can be used simultaneously. Our experiment and setup were approved by the Information Technology Risks Committee of our university.

\subsection{VM Hardening}
An issue that can arise from testing malicious software in a sandbox, using virtual machines, is the detection of the environment by the malware. Some malware have been shown to use multiple techniques to identify when they are running in a sandbox, and subsequently terminate their malicious activity or delete themselves~\cite{chen2008towards}. They employ sandbox and virtual machine detection to prevent researchers and others from analyzing their software. 

Modifications have been made to the virtual machines to ensure more resilience to virtual environment and sandbox detection from malware. The Cuckoo sandbox environment provides a \emph{disguise} module, which modifies various registry keys identifying the machine as a virtual environment. Since these registry key values are hard-coded in the disguise module, we modified them slightly to evade any malware also trying to identify the Cuckoo sandbox.

We opted for KVM as a virtualization software, since detection methods appear to be more prevalent for VMware~\cite{ferrie2007attacks} and that QEMU/KVM leaves fewer traces in the operating system.

Our virtual machines have been tested using Paranoid Fish~\cite{ortega_pafish_2019}, a popular virtual machine and sandbox fingerprinting tool. In total, 50 out of 54 tests yielded success, with some tests failing due to our limited hardware. However, upon inspection of a subset of our samples, they did not employ most of these techniques, and the ones present were covered by our modifications.

\subsection{Choice of Features}
\label{features}
Next, we must establish which features identify our machine profile, and what may affect the downloader's behavior, i.e., which features might be targeted by cybercriminals. In this work, we use a \emph{black box} approach, i.e., we do not reverse engineer downloader software to establish what it searches on the machine, but rather change various features of our environment and observe possible changes in behavior. We opted for this approach, considering there are thousands of different samples of downloaders in our dataset, and they might be modified frequently. Reverse engineering them would not be possible given the time and effort it requires. The malicious hosts serving downloaders might also analyze the machine features server side, in which case reverse engineering the samples would at best only provide partial information. 

We also opt to change features of our machines instead of monitoring Windows API calls made by the downloader, in order to limit our interactions within the VM as much as possible. This is to prevent the detection of the research environment by the malicious software, and to avoid impeding its execution in any way.

The features of our machines might notably be used in PPI networks, in order to decide which payload to install on a machine. Previous research~\cite{thomas2016investigating} has analyzed PPI networks and what information adware downloaders send to their control servers. In their analysis of 4 PPIs, they observed the following features being sent by the downloaders: 
\begin{itemize}
	\item The operating system and service pack version
	\item The Web browsers installed, along with their version
	\item The IP address
	\item Potentially unique identifiers, such as the MAC address
\end{itemize}

Although these PPIs are not necessarily similar to PPIs sending malware, using these features for our experiments will provide a basis to follow. 

Firstly, we test multiple OSes, namely Windows XP, Windows 7 and Windows 10. Windows was chosen given that it is the most common desktop OS\footnote{http://gs.statcounter.com/os-market-share/desktop/worldwide/\#monthly-201803-201803-bar}. Windows 10 and Windows 7 were used since they consisted in the most popular desktop OSes at the time, while Windows XP was chosen given that it is not supported anymore by Microsoft but still has users running it worldwide.

We used a VPN to change our location to a specific country. In order to establish the set of countries we wish to test, we identified the countries with the most Internet users\footnote{https://www.internetworldstats.com/top20.htm}. We could not obtain a VPN access in Japan, and thus chose to test Korea instead. We also included Iran in order to have at least one country from the Middle East. We could not obtain a VPN located in China, and instead used Hong Kong as the location. Our intuition is that, at the time of our experiments, malicious downloaders targeting China might also target Hong Kong. In short, our list of countries consists of: Hong Kong, Iran, the United States, Brazil, Nigeria, Korea, Russia, Germany, Mexico, Bangladesh.

Additionally, we also add features associated with the web history and the profile of the user. In the past, some malware has been shown to look at the keyboard or display language of a system when executed~\cite{mcafee2016banload, eset2009countryevade}. Thus, we opted to also include these features. The keyboard layouts and the display languages chosen were the top 10 languages of Internet users worldwide\footnote{https://www.internetworldstats.com/stats7.htm}. Thus, the languages chosen are: English, Chinese, Spanish, Arabic, Portuguese, Indonesian, French, Japanese, Russian, German. The keyboard languages and display languages are set by changing the corresponding Windows registry keys.

Finally, we identified nine Web browser session profiles and browser history profiles to test, based on Alexa's categories and their list of most popular Web sites\footnote{https://www.alexa.com/topsites/category}. To create browser Web sessions and a browser history, we open in the VM's browser the top ten most popular Web sites of the current category being tested, and wait for them to load before beginning the experiment. Our categories are the following: Business, Games, Health, Kids and teens, Men, News, Social networks, Sports, Women.

A summary of our features is described in Table~\ref{tab:features}.

\begin{table}[]
	\begin{center}
	\vspace*{-2mm}
	\begin{tabular}{lc}
	\hline
	\multicolumn{1}{|l|}{\textbf{Category}}         & \multicolumn{1}{c|}{\textbf{Feature}}                	\\ \hline
	\multicolumn{1}{|l|}{\multirow{3}{*}{User features}} & \multicolumn{1}{c|}{Country}               \\
	\multicolumn{1}{|l|}{}                  & \multicolumn{1}{c|}{Browser History}                   \\
	\multicolumn{1}{|l|}{}                  & \multicolumn{1}{c|}{Web session (browser cookies)} \\ \hline
	\multicolumn{1}{|l|}{\multirow{4}{*}{Configuration features}}        & \multicolumn{1}{c|}{IP address} \\
	\multicolumn{1}{|l|}{}                  & \multicolumn{1}{c|}{Windows OS version}                     \\
	\multicolumn{1}{|l|}{}                  & \multicolumn{1}{c|}{Display language}                              
\\
	\multicolumn{1}{|l|}{}              	& \multicolumn{1}{c|}{Keyboard language}                               \\ \hline
	\end{tabular}
	\caption{Tested features of the victim machine}
	\label{tab:features}
	\end{center}
	\vspace*{-7mm}
\end{table}

Other user features, such as the age, gender and income, while being relevant, are not always possible to identify on a machine, and thus, will not be considered in our experiments.

\subsection{Downloader Experiment}
\label{metho_downloaders}

Our second research objective is to establish the correlation between user profiles and the malicious payload. Our experiments employ our previously defined environment to run downloader samples.

The dataset, i.e., the downloader samples, is provided on a daily basis by our security vendor partner ESET, an antivirus software company. Each downloader sample is run, when received, on a set of VMs with profiles testing each of the various features. Any duplicate downloader received is discarded. After each set of tests of a downloader, our analyzer establishes what payload was downloaded on the victim machine. 
Our experimental design is as follows:

Two downloader families are tested, each of them for a period of 15 minutes at a time, so that any execution delay in the sample doesn't impact the test, and our profile set up has time to execute in the VM. Up to 10 samples of each family are tested per day, according to the availability in our data stream. We aim at testing the most current and widespread malicious downloaders, that are known to have dropped multiple types of malware. 

Our initial list of families consisted of Waski, Zurgop, Pliskal, Wauchos, Nymaim, Tovkater, Banload and Emotet, which all have the most mentions in security vendor blogs and the research literature, at the time of the beginning of our experiments. However, we could not obtain enough samples of Wauchos, Nymaim and Pliskal to consistently test these three families. While it is not clear why no samples of Pliskal were available, it is likely that Wauchos samples were less prevalent following a large disruptive operation by law enforcement authorities worldwide in late 2017~\cite{eset2018wauchos}. As for Nymaim, the lack of available samples might be explained by the fact that the family is older and less active in 2018. We ran some preliminary tests, where samples of each of the remaining families were run through our whole framework. After running these experiments for a week, we noted every family that made at least one successful HTTP connection to an external server, i.e., that had at least one online C\&C server. In the end, the only families to have at least one online C\&C server were Tovkater and Banload.

In total, we have 42 feature variations to test. To obtain a large enough set of tested samples, the variations have been tested each day for a period of 12 months, and the test variations have been executed each day at different times in order to limit a potential bias due to the time of execution. We can establish the number of VMs needed for this experimental design as follows:

\begin{definition}
The number of VMs needed can be calculated using $V = d * s * k * t / p$, where $d$ is the number of downloader families, $s$ is the number of samples of a family, $k=n*l$ is the number of feature variations, $t$ is the time needed to run each sample, $p$ is the total minutes in a day, and $V$ is the number of VMs needed to run the tests. 
\end{definition}

For our experiments, 42 feature variations are tested, with a running time of 15 minutes, where a day consists of 1440 minutes, i.e., 24 hours. A full factorial experiment is not possible, given that it would produce 27,000 feature variations, thus requiring 5,625 VMs. We opted to establish default values for each feature, where only a selected feature to test would be modified and tested. 
 
Our default profile consists of the most popular feature for each category, namely the United States as the country, English as the keyboard layout and the display language, and \emph{social networks} as the browser session.

Thus, using our formula, a minimum of 9 VMs are required to run our experiments each day. 

\subsection{Labeling}
\label{labeling}
Our final dataset is a list of instances of multiple features of our VMs associated with the data extracted from the execution of a malicious downloader in a sandbox. 

To test our hypothesis, we label each dataset entry with the family of the downloaded payloads from the execution of the malicious downloader. 

Each payload is tested on VirusTotal to establish if it is malicious, and if so, its family. Each sample is scanned a month after the end of our experiments, in order to take into account labels that change over time~\cite{zhu2020measuring}. In order to filter possible false positives, we only consider malware with at least 5 positive reports in VirusTotal to be malicious, following previous research establishing the ideal threshold at between 2 and 15~\cite{zhu2020measuring}.

Establishing the family of a malware sample, particularly a fresh new sample, is a difficult task that is still the subject of current research. Avclass~\cite{sebastian2016avclass} is a tool that aims at identifying malicious families through the various entries of a single VirusTotal report. While this tool works well when identifying large generic families, it proved unable to correctly differentiate our various families, labeling our samples as either a singleton (without a family), or all in one unique family. 

A number of other approaches aim at clustering malware into families~\cite{perdisci2012vamo, chen2018malcommunity, de2018efficient}. Unfortunately, works such as these are either tested on a specific subset of malicious families, or consist in a proof of concept without any published tool. Thus, we implement our own approach to identify malware families, for our particular setup.

Firstly, we inspect a subset of our payloads' VirusTotal reports, in order to identify the predominant labels for various payloads. These labels often consist in a generic identification, when the family cannot be correctly identified by the antivirus, e.g., a signature such as \emph{Trojan.Generic.XYZ}. These labels introduce noise into our data and are thus discarded. From this, we obtain a set of label families for each payload. The labels differ between antivirus for various reasons: different antivirus software might have different family names for the same malware, some might have more fine-grained family identification than others, and some antivirus software might correctly identify a malicious software but incorrectly identify its family. Previous research has identified that many AV signatures are incomplete or inconsistent~\cite{bailey2007automated}. Thus, we initially establish a subset of precise family names used by antivirus software through our manual VirusTotal reports analysis, and we then label the malicious payloads with the most occurring label. If no label from our list is present, we default to the majority label found, similar to Avclass. 

Our labeling approach simply aims at clustering similar samples together, so as to establish if a cluster behaves differently for a machine profile. Thus, our clustering follows the naming of antivirus companies only as a means of differentiating samples. While previous research demonstrated ways to identify malware by its behavior~\cite{rieck2011automatic, cho2014malware, wagener2008malware, bailey2007automated}, we limit our interactions inside our VMs to the minimum, and thus, do not make an in-depth analysis of our malicious downloaders' behavior to construct similarity profiles.

The following are our identified families and are used as a label when they are the most occurring family: Eldorado, Adware, KillAV, Psyme, Banload, InstallMonster.  

\subsection{Time Series Analysis}
In order to establish changes in the behavior of our downloader samples, we build time series out of our experiment data. The number of dropped payloads, for one particular family, as established in Section~\ref{labeling}, is extracted from our results for every week of our experiments. This is repeated for every feature tested as to identify all changes of behavior through time for our various downloader families. 

To test whether a change has happened in a time series, we employ changepoint analysis. Since our data consists of time series where we aim at identifying a change of behavior, we opted to use changepoint analysis to extract the various changepoints of each value of a feature, and thus, identify if one or more value of a feature changes at a different time than the others. Changepoint analysis is the detection of a change in the distribution of data in a time series, and it establishes the precise time at which a change occurs. Every point in time where a change occurs consists in a \emph{changepoint}. Multiple algorithms have been established in the past to improve the detection of changepoints~\cite{guralnik1999event, kawahara2009change, liu2013change, jandhyala2013inference}. Changepoint analysis has been employed in previous research in computer security, although mainly in network intrusion and anomaly detection~\cite{tartakovsky2006novel, tartakovsky2012efficient}. 
To execute our changepoint analysis, we've opted to use the Ruptures Python library~\cite{truong2018ruptures}, which implements a multitude of changepoint algorithms.

Finally, for every downloader family and feature type, we compute the ratio of infections to total runs for each day of our experiments, and perform a one-way analysis of variance (ANOVA) of the means of the ratio of infections for each feature and each type of payload. 

For this, we keep every day with at least one infection for one value of a feature and use this dataset for our analysis. We use the ratio of infections instead of the number of infections to compensate for any imbalance in the number of downloader runs per feature. While our \emph{Profiler} manages the queue of experiments and equally distributes the features to test, some machines in our cluster were the victim of hardware issues and outages during the one-year experiment, and were sporadically offline. 

The statistical analysis is performed using SigmaPlot (Systat). The data is tested for normality using the Shapiro-Wilk test, and tested for homoscedasticity using the Kolmogorov-Smirnov test. All the tests are two-sided. Following this, we run a one-way ANOVA on ranks of our values, since the distribution of the ratio of infection is not a normal distribution. We employ the Kruskal-Wallis test~\cite{kruskal1952use}, which is a non-parametric method to determine the difference between the means of different groups.

\section{Results}
\label{results}
We employed our framework over a one-year period, from February 2018 to February 2019, and ran malicious downloaders on our predefined machine profiles. We collected data on these executions and established time series of infections for our various downloader families and payload families. 

\subsection{Dataset}
Our final downloader dataset consists of 1,526 unique samples, where 711 were identified as part of the Tovkater family, and 805 were identified as part of the Banload family. On average, each sample was run 100 times, when testing our profile features according to our experimental design. In total, we ran 151,189 tests inside our VMs through the Cuckoo framework, with 72,829 tests run on Tovkater samples and 78,360 on Banload samples. Of these, 18,975 resulted in a \emph{detonation} of the malicious downloader, i.e., these downloaders downloaded other malware payloads from the Internet. While most downloaders only fetched one payload per run, there were 258 that downloaded two payloads and 163 that downloaded three payloads. Multiple types of files were downloaded by the downloaders, depending on the initial sample and the needs of the C\&C. The file types identified by inspecting the file header of the downloaded payloads are: octet-stream, x-dosexec, vnd.ms-cab-compressed, vnd.openxmlformats-officedocument, vnd.ms-powerpoint(ppt/pptx), sqlite, x-shockwave-flash, zip, plain, html, x-msdos-batch, xml, ini file, vbscript, json, png, jpeg, gif, svg and x-wav.

The text, HTML and octet-stream payload files generally indicated a failure to correctly run a downloader. 

Other executions successfully detonated, and upon inspection, some began the infection with a batch script, a javascript page or a VBScript launching or downloading another malicious executable. A series of Banload runs had an initial batch file, which downloaded an executable, which in turn downloaded a malicious powerpoint file. Another type of multi-stage attack was observed for Banload downloaders, where they retrieved a Microsoft Office document, which then installed another malicious executable. In some cases, we observed an \emph{autoliker} chrome extension being downloaded, hinting at a possible use of the infected machine as a bot on social networks.

A number of ZIP files were downloaded by our downloaders. Upon closer inspection, these did not seem to be valid ZIP files, but did appear to contain a malicious executable.

Finally, to confirm the maliciousness of the retrieved payloads, we tested each one on VirusTotal, and ended up with 3,608 malicious payloads.  

\subsection{Labeling}
We employed the technique described in section~\ref{labeling} to label all of our collected instances that resulted in a detonation. 

These samples had at least 5 positive detections in VirusTotal, and 1298 samples had a majority of positive detections.

In total, our collected payloads consist of 75 instances of KillAV, 134 instances of Psyme, 318 instances of Banload, 346 instances of Eldorado, 1,121 instances of Adware, and finally 1,580 instances of InstallMonster. As can be observed, the majority of our downloaded payloads were labeled as Adware or InstallMonster. 

Previous research identified that PUP and malware share infrastructure~\cite{ife2019waves,kwon2016catching}, which explains the presence of Adware samples in the retrieved payloads of our downloaders.

In order to identify if our respective downloader families are used in a PPI distribution model, we computed the number of distinct payload families served. On average, Banload downloader binaries each served 1.5 families, while Tovkater binaries served 2.06 distinct families. In total, the following families were seen in dropped payloads of Banload: KillAV, Eldorado, Adware, Zpevdo, Installmonster, Banload, Psyme and Wisdomeyes. For our second downloader family, Tovkater, the following families were seen in malicious payloads: Adware, Zpevdo, Eldorado and Installmonster.

\subsection{Infections Through Time}

When observing the rate of infections for each of our labeled families, according to our various features, we can establish where a peak of infections occurs for a specific feature and a specific malware family. We applied changepoint analysis to our various time series, and noted the feature values for which these vary. Our results also differ when testing our two downloader families. As the tests were running every day and the changepoint analysis needed less granularity, we created four bins for each month (on the 4th, 12th, 20th, and 28th) grouping the activities of the 7-8 days around them. Additionally, we ran a one-way ANOVA of the ratio of infection per day to identify if the features identified as statistically different match with our features identified in the changepoint analysis.

All data in the figures are presented as the mean + Standard Error of the Mean (SEM). The asterisks in our figures denote the statistical significance for specified tests, chosen as $* = p < 0.05$, following a previous study \cite{lalonde2013clinical}. 

Next, we will show the results returned by the application of the changepoint analysis on the different features, along with the results of our analysis of variance.

One of the most impactful features in our tests was, unsurprisingly, the operating system used in the VM running the downloader executable. 

When running Windows XP, we obtained less than half the number of infections of other OSes. The majority of the Windows 7 infections belong to the InstallMonster family, whereas Windows 10 is mainly a victim of Adware.

\begin{figure}[!h] 
	\centering 
	\begin{minipage}[t]{0.45\textwidth}
		\centering 
		\vspace*{-5mm}
		\includegraphics[width=1\linewidth]{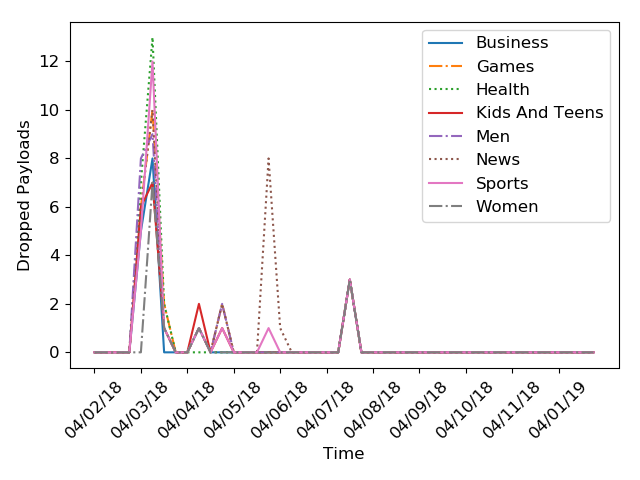} 
		\caption{The number of dropped Adware payloads over time according to the browser profile} 
		\label{fig:adware_browser}
	\end{minipage}
	\begin{minipage}[t]{0.45\textwidth}
		\centering 
		\includegraphics[width=0.9\linewidth]{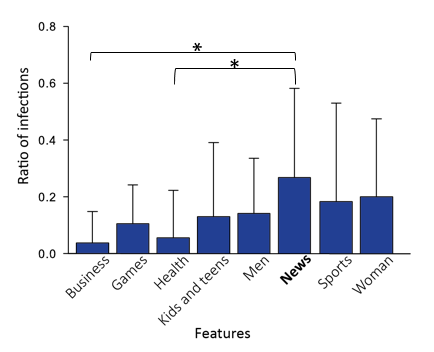} 
		\caption{The average ratio of infections of Tovkater with Adware payloads over time according to the browser profile. Significantly different averages are marked in bold.} 
		\label{fig:adware_browser_anova}
		\vspace*{-5mm}
	\end{minipage} 
\end{figure}

The Windows version used in the VM is not the only feature highlighted by our analysis. Firstly, Figure~\ref{fig:adware_browser} highlights infections that downloaded Adware as a payload while testing our browser profile feature and running the Tovkater downloader. As can be observed, there is an increase to 8 infections with solely a \emph{news} browser profile, where the closest other browser profile is \emph{sports} with 2 infections. Our changepoint analysis identified four common changepoints among the different browser profiles: 
\begin{enumerate}
	\item The 4th of March 2018: The start of our experiments and an increase in the number of infections
	\item The 12th of April 2018: The moment of a second sudden peak of infections
	\item The 28th of June 2018: The end of the first infection pause mentioned earlier
	\item The 4th of August 2018: The separation between the activity in the summer and the decrease following it
\end{enumerate}

The algorithm identified another relevant point in its analysis: the 20th of May 2018. The changepoint analysis highlighted that the \emph{news} browser profile was showing different activities from the other ones. In fact, this profile has seen more detonations, resulting in a higher number of payloads downloaded in this configuration.

We ran a one-way ANOVA to compare the means of the ratio of infections by Adware for different browser profiles when running Tovkater, for the time period with the most activity: from March 6 2018 to June 6 2018. One browser profile was identified as being different: the \emph{news} browser profile results were identified as statistically higher than the \emph{business} and \emph{health} browser profiles with a p-value\textless 0.05. Indeed, the news browser profile shows an average ratio of infection more than two times higher than infections with the health or business profile, which confirms that our machines with the news browser profile received more Adware payloads. The means of the ratio of infections for each browser profile are identified in Figure~\ref{fig:adware_browser_anova}.

Secondly, we tested our various display languages, and observed some changepoints differences. However, when running an ANOVA on the means of the ratio of infection, no feature was identified as statistically different.

\begin{figure}[h] 
	\centering 
	\begin{minipage}[t]{0.45\textwidth}
		\centering 
		\vspace*{-4mm}
		\includegraphics[width=1\linewidth]{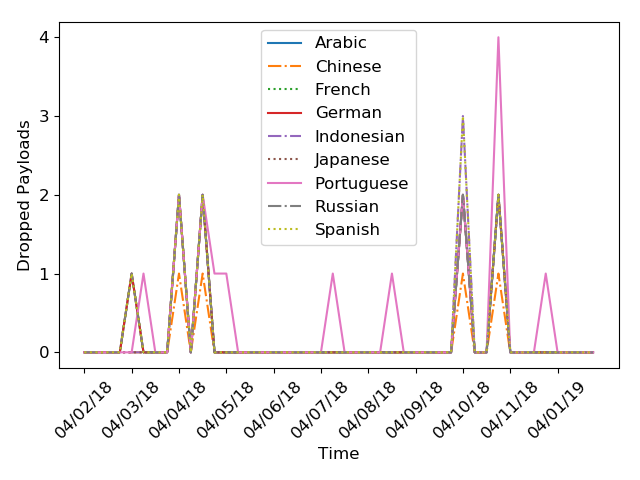} 
		\caption{The number of dropped Banload payloads over time according to the keyboard layout} 
		\label{fig:banload_keyboard}
	\end{minipage}
	\begin{minipage}[t]{0.45\textwidth}
		\centering 
		\includegraphics[width=0.9\linewidth]{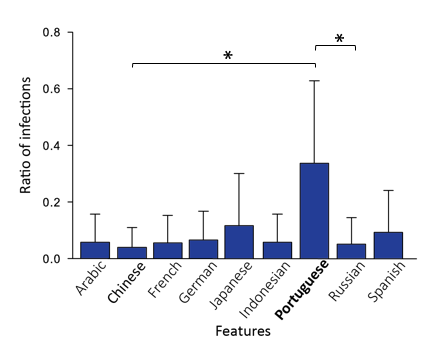} 
		\caption{The average ratio of infections of Banload dropping Banload payloads over time according to the keyboard layout. Significantly different averages are in bold.} 
		\label{fig:banload_keyboard_anova}
		\vspace*{-3mm}
	\end{minipage} 
\end{figure}

Another interesting activity was observed when testing the different keyboard layouts with the Banload downloader, as shown in Figure~\ref{fig:banload_keyboard}. Specifically for dropped payloads part of the Banload family, more activity was seen over time when the keyboard layout consisted in Portuguese. This result can be further confirmed by analyzing online activity of the Banload downloader\footnote{\url{https://www.virusradar.com/en/Win32_TrojanDownloader.Banload/map}}\footnote{\url{https://securityboulevard.com/2019/05/cybercrime-groups-behind-banload-banking-malware-implement-new-techniques/}}, which is most active in Brazil, a country with Portuguese as its official language. Moreover, another keyboard layout with which the downloader is behaving differently is Chinese, where there are fewer detonations than with our other keyboard layouts. Our changepoint analysis identified four changepoints for all our keyboard layout values except Chinese:
\begin{enumerate}
	\item The 4th of March 2018: The start of our experiments
	\item The 20th of May 2018: A drop in the number of infections
	\item The 12th of September 2018: Infections suddenly increased
	\item The 28th of December 2018: The infections' final drop
\end{enumerate}

It is worth noting that there is a smaller number of infections than for our previous results. The Chinese keyboard language does not have any changepoints identified, which highlights the fact that VMs with a Chinese keyboard language are less infected. The changepoint analysis on the data related to the Portuguese keyboard language, however, has highlighted two additional moments in the timeline, the 28th of June 2018 and the 20th of October 2018. These moments correspond to an increased number of infections which is not happening when using the other layouts.

We also ran an ANOVA on the ratio of Banload infections for keyboard layouts when running the Banload downloader, and found one of them to be statistically different from March 20 2018 to June 6 2018. The Portuguese keyboard layout was found to be statistically higher than the Chinese and Russian keyboard layouts, with a p-value\textless 0.05. 

The means of the ratio of infection of each keyboard layout is shown in Figure~\ref{fig:banload_keyboard_anova}, where we can observe that the Portuguese keyboard layout obtains more than twice the ratio of infections of the Russian and Chinese keyboard layouts. 

\begin{figure}[h]
 \centering
 \vspace*{-4mm}
 \includegraphics[width=0.85\linewidth]{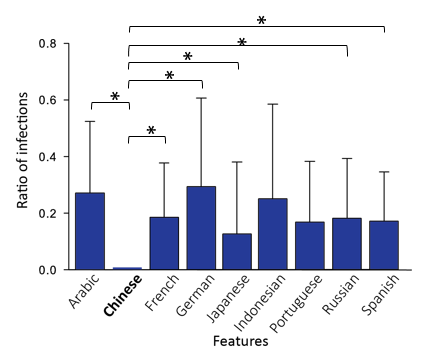}
 \caption{The average ratio of infections of Tovkater dropping Adware payloads over time according to the keyboard layout. Significantly different averages are in bold.}
 \label{fig:tovkater_keyboard_anova}
 \vspace*{-4mm}
\end{figure}

The Chinese keyboard layout appears to be the victim of less Adware infections when running Tovkater downloader samples as well. In fact, no Adware infection was registered at all for this keyboard layout. We ran an ANOVA on this feature and identified the Chinese keyboard layout to be different than almost all other layouts: the ratio of infection is statistically lower than for the Arabic, German, Russian, Japanese, French and Spanish keyboard layouts with a p-value\textless 0.05.
The means of the ratio of infections per keyboard layout can be observed in Figure~\ref{fig:tovkater_keyboard_anova}.

Finally, while changepoints were identified for some peaks of infections for machines in different locations, we ran an ANOVA of the ratio of infections and did not find any statistically different locations.

\section{Discussion}

Through our experiments, we aimed at reverse engineering the targeting choice of criminals by establishing the link between a malicious downloader and the profile of the victim machine. Our approach did not look at the code of malicious samples, but empirically determined this link by setting up different VM profiles and analyzing whether samples were reacting to different profiles in different ways.
In Section \ref{results}, we described the findings of our experiments, i.e., which profiles attracted a certain family of malware, and to what degree, with the number of infections. The changepoint analysis highlighted the important moments during the 12 months of experiments, and an analysis of variance identified if the average ratio of infection for a given feature value was significantly different than other values.

The highest number of malicious payloads was experienced on machines running Windows 10, while Windows XP resulted in the least infections. The popularity of Windows 10 among these malware families is particularly interesting as the OS has become more prevalent than Windows 7 only at the beginning of 2019\footnote{\url{https://www.theverge.com/2019/1/2/18164916/microsoft-windows-10-market-share-passes-windows-7-statistics}}.
Windows XP is not supported anymore by Microsoft, and is thus less secure than other modern OSes, however, these results show how it is becoming less relevant to the cybercriminals' purposes. One explanation might be that malware is targeting the most used OSes, and as such avoid Windows XP. Additionally, droppers downloaded different families of malware depending on the OS. Windows XP and Windows 7 were mainly infected with InstallMonster payloads, while Adware payloads were used to target Windows 10 VMs.

Similar to previous work, our analysis has highlighted how the OS is a feature taken into consideration by criminals. 
However, our results also presented a new phenomenon, i.e., that the browser profile and keyboard layout can have an impact on what payload a downloader downloads. 

In the previous section we mentioned that at one point, Tovkater samples focused on the download of Adware payloads when observing a \emph{news} browser profile. 

Indeed, users visiting news websites might be more susceptible to adware, or correspond more closely to the target demographic of the malicious actors. We also noted an increase in Banload downloaders downloading additional Banload payloads when running a machine with a Portuguese keyboard layout, particularly compared to the Chinese and Portuguese keyboard layouts.

This is corroborated by news articles showing that a Banload campaign has run since May 2018\footnote{\url{https://securityboulevard.com/2019/05/cybercrime-groups-behind-Banload-banking-malware-implement-new-techniques/}}. Malicious actors appear to target Portuguese-speaking countries through the Banload downloader.

The Chinese keyboard layout also did not receive any Adware infections when running Tovkater samples, further confirming that this keyboard layout is less targeted by Adware samples. One explanation for this phenomenon can be that malicious actors avoid targeting their country of residence. In can also be that laws around adware infections are stricter in China, or that the downloader operators fear harsher penalties. Finally, it can simply be that Chinese-speaking countries are less attractive to these malicious actors, due to there being fewer profits to be made.

These findings highlight how crucial it is to consider the context in which a malicious downloader is executed when trying to detonate it and observe its behavior. One of the key issues security researchers face when analyzing malware is effectively executing them in a research environment in order to identify their malicious behavior and build methods to detect and mitigate them. Our results show how identifying important profile features for a malicious downloader can not only have an impact on the number of downloaded payloads, but also on the type of downloaded payloads as well. 

\paragraph{Limitations}
While our results have shown how Tovkater and Banload downloaders behave differently given various machine profiles, the downloader samples used in our experiments were provided by our antivirus partner, and as such, are only samples that could be detected and retrieved by them. Thus, these samples might not necessarily be representative of the general ecosystem of malicious downloaders. We also have limited information regarding the source and context from which originated the downloader samples. 

Another limitation is that we tested each feature independently, in order to clearly identify if one feature impacted the execution of a downloader. We did not possess enough resources to test combinations of features, such as a matching keyboard and display language. However, while this setup might have negatively impacted the number of noteworthy detonations of malicious downloaders, we still obtained a number of results solely with our single feature tests.
\section{Conclusion}

In this work, we aimed at reverse engineering the targeting choice of cybercriminals acting through PPI networks, by establishing if a link between the features of a machine and the payload(s) downloaded by a malicious downloader exists. We successfully built an automated sandboxed environment framework, which is capable of changing the configuration of a VM for a specific run of a malicious downloader executable. Using this setup, we ran 151,189 tests on Tovkater and Banload downloader families. Of these, 18,975 resulted in the download of at least one additional payload. With the use of changepoint analysis applied on time series of our infections and ANOVAs of the ratio of infection per day, we identified different malicious payloads downloaded depending on the operating system.

More notably, we showed that malicious downloader families download different payloads depending on the browser history and the keyboard layout of the machine, highlighting the targeting choices of cybercriminals when infecting victims through downloaders. 

Our findings show that an effective setup to analyze malicious downloaders should consider the features of the virtual machines in order to obtain better rates of detonation and to gather a larger array of malware families. 

\begin{acks}
We would like to thank ESET for their technical and financial support, as well as for providing data for our experiments. We would also like to thank \`Eve Honor\'e for her help in the statistical analysis. This work was partially supported by the NSF under Grant CNS-2127232.
\end{acks}

\bibliographystyle{ACM-Reference-Format}
\bibliography{refs}

\end{document}